\documentclass[aps,pra,showpacs,groupedaddress,superscriptaddress,twocolumn]{revtex4}
\usepackage[dvips]{graphics,color}
\usepackage{graphicx}
\usepackage{latexsym}

\newcommand{\be}{\begin{equation}}
\newcommand{\ee}{\end{equation}}
\newcommand{\br}{\begin{eqnarray}}
\newcommand{\er}{\end{eqnarray}}

\newcommand{\bd}{\begin{displaymath}}
\newcommand{\ed}{\end{displaymath}}

\newcommand{\bfig}{\begin{figure}}
\newcommand{\efig}{\end{figure}}
\input{tcilatex}
\begin{document}

\title{Engineering Quantum Jump Superoperators for Single Photon Detectors}
\author{A. V. Dodonov}
\email{adodonov@df.ufscar.br}
\author{S. S. Mizrahi}
\email{salomon@df.ufscar.br}
\affiliation{Departamento de F\'{\i}sica, CCET, Universidade Federal de S\~{a}o Carlos,
Via Washington Luiz km 235, 13565-905, S\~ao Carlos, SP, Brazil}
\author{V. V. Dodonov}
\email{vdodonov@fis.unb.br}
\affiliation{Instituto de F\'{\i}sica, Universidade de Bras\'{\i}lia, PO Box 04455,
70910-900, Bras\'{\i}lia, DF, Brazil}
\date{\today }

\begin{abstract}
We study the back-action of a single photon detector on the
electromagnetic field upon a photodetection by considering a
microscopic model in which the detector is constituted of a sensor
and an amplification mechanism. Using the quantum trajectories
approach we determine the Quantum Jump Superoperator (QJS) that
describes the action of the detector on the field state
immediately after the photocount. The resulting QJS consists of
two parts: the bright counts term, representing the real
photoabsorptions, and the dark counts term, representing the
amplification of intrinsic excitations inside the detector. First
we compare our results for the counting rates to experimental
data, showing a good agreement. Then we point out that by
modifying the field frequency one can engineer the form of QJS,
obtaining the QJS's proposed previously in an \emph{ad hoc}
manner.
\end{abstract}

\maketitle

\section{Introduction}

Single photon detectors (SPD) represent the ultimate sensitivity limit for
quantum photodetectors, and many quantum optics and quantum information
applications are based on its existence \cite{JMO-04}. Nowadays, there are
several available types of SPD's sensitive to different light wavelengths
and with a varying range of quantum efficiencies \cite%
{QD,APD1,Verevkin,Kitay,SSPD,APD2,photo,TES,VLPC,kor,Karve}. Among
many applications, SPD's are the main ingredient in the situations
where one measures the electromagnetic field (EM) with few photons
enclosed in a cavity, the photons being counted one by one.
Theoretical treatment of such a scheme, known as \emph{continuous
photodetection model} (CPM), was proposed by Srinivas and Davies
(SD) in 1981 \cite{SD} and did find many applications since then
\cite{appl,appl1,appl2}.

In SD scheme, in each infinitesimal time interval the photodetector has only
two possible outcomes: either a single photon is detected (`click' of the
detector), or it is not. In both cases the state of the field changes as
time goes on: for a click, the field loses one photon and suffers a \emph{%
Quantum Jump}; for no-click the field state modifies continuously
and non-unitarily due to the monitoring of the detector
\cite{uedas,DMD-JOB05}. Thus, besides allowing determination of
the statistical properties of the EM field through photocounts
statistics, the detector also exerts a back-action on the field by
which the outcome of the measurement modifies the cavity field
state \cite{ueda1}. This phenomenon was widely used for different
theoretical proposals, e.g. for changing the field statistics from
sub-Poissonian to super-Poissonian \cite{Ban}, for controlling the
entanglement between two field modes \cite{appl1} or inducing spin
squeezing in a cavity \cite{appl2}. However, no experimental
verification of the CPM has been made until now, though nowadays
it is quite realistic to make simple photocounting experiments for
testing the theory, provided one takes into account inevitable
losses present in the experiment and include them into the model
\cite{nonideal,prepar}.

Here we study how the field state after the click depends on
detector's parameters. It is worth considering how SPD's actually
operate: despite technical and structural differences associated
with every kind of detector, the photodetection process is based
on the same principle: the `sensor' initially set in a `ground
state' interacts with the field and is likely to absorb a photon,
doing a transition to the `excited state'. After some time the
sensor decays back to the ground state, emitting a photoelectron
that triggers the `amplification mechanism' (AM) of the SPD (e.g.,
by an avalanche process), producing a pulse of macroscopic
electric current or voltage, that originates a registered click of
the detector, representing one count. Besides, in real
photodetectors there is a phenomenon called \emph{dark counts}:
photoelectrons originated due to the intrinsic excitations inside
the AM, and not due to the absorption of one photon from the
field. The influence of dark counts on the results of various
experiments was considered in
\cite{Duan01,Par01,Mir05,Boil05,Inver05}, and different schemes of
calculating dark count probability and single-photon quantum
efficiency were used in \cite{Panda,Wars02,Kang03}.

In the CPM, all the results concerning the photodetection process are
described by means of a single entity characterizing the photodetector ---
the Quantum Jump Superoperator (QJS) $\hat{J}$, that represents the
back-action of the detector on the field upon a single photodetection.
Immediately after the photodetection an initial field state described by the
statistical operator $\rho $ changes abruptly to $\rho ^{\prime }=\hat{J}%
\rho/{\mathrm{T}r}[ \hat{J}\rho] $, and the probability for registering one
count during the time interval $[t,t+\Delta t)$ is $\mathrm{Tr}[\hat{J}\rho
]\Delta t$, where $\Delta t$ is the time resolution of the detector. It is
supposed that $\Delta t$ is small compared to other characteristic time
scales and that QJS is time-independent, in an ideal case being given by
\begin{equation}
\hat{J}\rho \equiv \gamma _{O}\hat{O}\rho \hat{O}^{\dagger },
\end{equation}%
where $\hat{O}$ is a lowering operator responsible for subtracting one
photon from the field and $\gamma_O $ is roughly the counting rate \cite%
{DMD-JOB05} with dimension (time$^{-1}$).

Srinivas and Davies proposed \emph{ad hoc\/} $\hat{O}=\hat{a}$, where $\hat{a%
}$ is the bosonic lowering operator. In spite of having some inconsistencies
as noted by the authors themselves, this QJS has been widely used since
then. Recently, another QJS defined with $\hat{O}=\hat{E}_{-}\equiv \left(
\hat{n}+1\right) ^{-1/2}\hat{a}$ (where $\hat{n}\equiv \hat{a}^{\dagger }%
\hat{a}$) was proposed, also \emph{ad hoc\/}, in \cite{benaryeh,OMD-JOB} --
we named it `E-model' for simplicity. The differences between these two
QJS's were studied in \cite{DMD-JOB05}, showing that the inconsistencies of
the SD-model are indeed eliminated.

In \cite{QJS} we have proposed a microscopic model (some other
models were considered in \cite{War,Rohde}) for the field-detector
interaction, where we showed that both QJS's proposed \emph{ad hoc}
are particular cases that can be derived from a general
time-dependent `\emph{transition}' superoperator. However, there we
considered a simplified model of the detector at `zero temperature',
by assuming that there were no intrinsic excitations inside the
photodetector and that the sensor was at exact resonance with the
field mode. Here we relax these conditions, being less stringent, we
take into account the effects implied by a `non-zero temperature'
detector possessing intrinsic excitations and allow a detuning
between the field and sensor frequencies. Moreover, we attribute
numerical values for the model parameters in order to reproduce
experimental data both qualitatively and quantitatively. We show
that the dark counts appear naturally in our model, and comparing
the predicted counting rates and Signal-to-Noise ratio with
experimental data, we get a good qualitative and quantitative
agreement. Furthermore, we demonstrate that the actual expression
for the QJS should be incremented as (cf. Ref. \cite{Duan01})
\begin{equation}
\hat{J}\rho =\gamma _{O}\hat{O}\rho \hat{O}^{\dagger }+\gamma _{D}\hat{D}%
\rho \hat{D}^{\dagger },
\end{equation}%
where $\gamma _{D}$ is roughly the dark counts rate and the
operator $\hat{D} $ describes the back-action of the detector on
the field due to dark counts. Finally, we point out that by simply
modifying the field frequency we are able to engineer the QJS,
thus obtaining either the SD-model or E-model in specific regimes.

The paper is organized as follows. In Sec. II we model the sensor
of the photodetector as a 2-level system, according to the well
known Jaynes--Cummings model; taking into account the effects of
the sensor--AM coupling, we obtain an explicit form of the
transition superoperator, from which we derive a general
expression for the QJS. In Sec. III we compare our results for
counting rates with experimental data and obtain specific
expressions of the QJS for different field wavelengths. The
section IV contains the summary and conclusions.

\section{Modelling the photodetector}

We assume the SPD as being constituted of two parts: the sensor
and the AM. The sensor is modelled as a two-level quantum object
with resonant frequency $\omega _{0}$, interacting with the
mono-modal EM field with frequency $\omega $ (for multi-modal
field one should just consider a frequencies distribution). It has
a ground and an excited states $|g\rangle $ and $|e\rangle $
(before and after the photoabsorptions), so we describe it by the
usual Jaynes--Cummings Hamiltonian \cite{JCM} (for its
applicability see, e.g. \cite{klimov,crisp} and references
therein)
\begin{equation}
H_{0}=\frac{1}{2}\omega _{0}\sigma _{0}+\omega {\hat{n}}+g\hat{a}\sigma
_{+}+g^{\ast }\hat{a}^{\dagger }\sigma _{-} ,  \label{JCH}
\end{equation}%
where $g$ is the sensor-field coupling constant, {and }the sensor
operators are $\sigma _{0}=|e\rangle \langle e|-|g\rangle \langle g|,$ $%
\sigma _{+}=|e\rangle \langle g|$ and $\sigma _{-}=|g\rangle \langle e|$. We
assume $g$ being real, since only its absolute value enters the final
expressions.

Upon absorbing a photon the sensor initially in the ground state
jumps to the excited state and some time later it decays back
emitting a photoelectron into the AM. The AM is a complex
macroscopic structure that depends on the type of SPD; it
amplifies the photoelectron and originates some observable
macroscopic effect, giving rise to the clicks of the detector. In
order to describe general features of the AM independent of the
type of SPD, we model it as a thermal reservoir with a mean
intrinsic excitations number $\bar{n}$. Thus, the whole system
field--SPD is described by the effective `standard master
equation' \cite{carmichael}
\begin{eqnarray}
\dot{\rho}_{T} &=&\frac{1}{i}\left[ H_{0},\rho _{T}\right] -\gamma \overline{%
n}\left( \sigma _{-}\sigma _{+}\rho _{T}-2\sigma _{+}\rho _{T}\sigma
_{-}+\rho _{T}\sigma _{-}\sigma _{+}\right)  \nonumber  \label{eqmestra} \\
&-&\gamma \left( \overline{n}+1\right) \left( \sigma _{+}\sigma _{-}\rho
_{T}-2\sigma _{-}\rho _{T}\sigma _{+}+\rho _{T}\sigma _{+}\sigma _{-}\right)
,
\end{eqnarray}%
where $\gamma$ is the sensor-AM coupling constant (we do not
consider the field damping due to cavity losses).

According to CPM the trace of QJS gives the probability density $\mathrm{{p}%
(t)}$ for the photodetection, i.e. emission of a photoelectron at time $t$,
given that at time $t=0$ the detector-field system was in the state
\begin{equation}
\rho _{0}=|g\rangle \langle g|\otimes \rho ,  \label{rhho0}
\end{equation}%
where the field state is $\rho $. Microscopically this means that
in the time interval $(0,t)$ the sensor has undergone a transition
$|g\rangle
\rightarrow |e\rangle$ due to the absorption of one photon, and $\mathrm{{p}%
(t)\Delta t}$ is the probability for decaying back to the ground state
during the time interval $[t,t+\Delta t)$, emitting a photoelectron that
will lately originate one click. Following the quantum trajectories approach
\cite{carmichael}, $\mathrm{p}(t)$ is calculated from the master-equation (%
\ref{eqmestra}) by identifying the sensor decay superoperator
\begin{equation}
\hat{R}\rho _{0}=2\gamma \left( \overline{n}+1\right) \sigma _{-}\rho
_{0}\sigma _{+}  \label{decay}
\end{equation}%
describing the instantaneous $|e\rangle \rightarrow |g\rangle$ decay and the
consequent emission of a photoelectron. The no-decay superoperator $\hat{U}%
_{t}\rho _{0}=\rho _{U}(t)$ describes the non-unitary evolution of the
field--SPD system from $t=0$ to $t>0$ without emission of photoelectrons; $%
\rho _{U}(t)$ is the solution to the master equation (\ref{eqmestra})
without the decay term (\ref{decay}):
\begin{equation}
\frac{d}{dt}\rho _{U}=-i(H_{e}\rho _{U}-\rho _{U}H_{e}^{\dagger })+2\gamma
\overline{n}\sigma _{+}\rho _{U}\sigma _{-},  \label{rhoU11}
\end{equation}
where the effective non-Hermitian Hamiltonian is
\begin{equation}
H_{e}=\frac{\left( \omega _{0}-i\gamma \right) }{2}\sigma _{0}+\omega {\hat{n%
}}+g\hat{a}\sigma _{+}+g\hat{a}^{\dagger }\sigma _{-}-i\gamma (%
\overline{n}+\frac{1}{2}).  \label{rhoU12}
\end{equation}%
Thus the probability density for the observation of a photocount,
or a click, at time $t$ is equal to
\[
\mathrm{p}(t)=\mathrm{Tr}_{F-D}\left[ \hat{R}\hat{U}_{t}\rho _{0}\right] ,
\]
where $\hat{R}\hat{U}_{t}\rho_0$ represents an evolution of the
field-SPD system from initial state $\rho_0$ at time $t=0$ to the
time $t$ without any decay of the sensor and an instantaneous
decay at the time $t$. Moreover, taking the trace only of the
detector variables, one obtains the expression describing the
action of the detector on the field upon the click -- a
predecessor of QJS that we call \emph{transition superoperator}
\begin{equation}
\hat{\Xi}(t)\rho =\mathrm{Tr}_{D}\left[ \hat{R}\hat{U}_{t}\rho _{0}\right] ,
\label{transition}
\end{equation}%
from which the probability density for a count is $\mathrm{{p}(t)=%
\mathrm{Tr} \left[ \Xi (t)\rho\right] }$.

Thus, for obtaining the transition superoperator one should first
determine
the no-decay superoperator $\hat{U}_{t}$ and then evaluate Eq. (\ref%
{transition}) using the initial state (\ref{rhho0}). In order to solve Eq. (%
\ref{rhoU11}) we do the transformation
\begin{equation}
\rho _{U}=X_{t}\tilde{\rho}_{U}X_{t}^{\dagger },  \label{data11}
\end{equation}
where
\begin{equation}
X_{t}=\exp (-iH_{e}t),  \label{data12}
\end{equation}%
and obtain a simple equation for $\tilde{\rho}_{U}$
\begin{equation}
\frac{d}{dt}\tilde{\rho}_{U}=2\gamma \overline{n}\widetilde{\sigma }_{+}%
\tilde{\rho}_{U}\widetilde{\sigma }_{-},
\end{equation}
\begin{equation}
\widetilde{\sigma }_{+}=X_{-t}\sigma _{+}X_{t},\qquad \widetilde{\sigma }%
_{-}=\widetilde{\sigma }_{+}^{\dagger },
\end{equation}%
whose formal solution is
\begin{equation}
\tilde{\rho}_{U}(t)=\rho _{0}+2\gamma \overline{n}\int_{0}^{t}dt^{\prime }%
\widetilde{\sigma }_{+}(t^{\prime })\tilde{\rho}_{U}(t^{\prime })\widetilde{%
\sigma }_{-}(t^{\prime }).  \label{sasa}
\end{equation}%
Now one iterates Eq. (\ref{sasa}) and obtains a power expansion in terms of $%
\bar{n},$ substitutes the result into Eq. (\ref{data11}) and then evaluates
Eq. (\ref{transition}), finally obtaining
\begin{equation}
\hat{\Xi}(t)\rho =2bg\left( 1+\bar{n}\right) \sum_{l=0}^{\infty }\left(
2\gamma \bar{n}\right) ^{l}\hat{\Xi}_{l}(t)\rho ,  \label{ttt11}
\end{equation}
where for $l>0$
\begin{equation}
\hat{\Xi}_{l}(t)\rho =\int_{0}^{t}dt_{1}\cdots \int_{0}^{t_{l-1}}dt_{l}\hat{%
\theta}_{l}\rho \hat{\theta}_{l}^{\dagger }  \label{ttt12}
\end{equation}
\begin{equation}
\hat{\theta}_{l}(t,t_1,\cdots,t_l) =\langle e|X_{t}\widetilde{\sigma }%
_{+}(t_{1})\widetilde{\sigma }_{+}(t_{2})\cdots \widetilde{\sigma }%
_{+}(t_{l})|g\rangle .  \label{ttt13}
\end{equation}%
For $l=0$ the integrals and the terms $\widetilde{\sigma }_{+}$ should be
dropped out:
\begin{equation}\label{ccc}
\hat{\Xi}_{0}(t)\rho = \hat{\theta}_{0}\rho
\hat{\theta}_{0}^{\dagger },\quad \hat{\theta}_{0}(t) =\langle
e|X_{t}|g\rangle .
\end{equation}

After some algebraic manipulations \cite{Sten,Cress96} we get for (\ref%
{data12})%
\begin{eqnarray}
X_{t} &=&\exp \left[ -\gamma t\left( \overline{n}+1/2\right) -i\omega {\hat{n%
}}t\right]  \label{xxx} \\
&\times &\left\{ \chi _{\hat{n}+1}(t)|e\rangle \langle e|+\chi _{\hat{n}%
}(-t)|g\rangle \langle g|\right.  \nonumber \\
&&-ie^{-i\omega t/2}S_{\hat{n}+1}(t)\hat{a}\sigma _{+}-ie^{i\omega t/2}S_{%
\hat{n}}(t)\hat{a}^{\dagger }\sigma _{-}\},  \nonumber
\end{eqnarray}%
where%
\begin{equation}\label{ggg}
C_{\hat{n}}(t) =\cos \left( \gamma tB{_{\hat{n}}/b}\right) ,\text{ }S_{\hat{n%
}}(t)=\sin \left( \gamma tB{_{\hat{n}}}/b\right) /{B}_{\hat{n}},
\end{equation}
\begin{equation}
\chi _{\hat{n}} =e^{-i\omega t/2}\left[ C_{\hat{n}}(t)-i\delta S_{\hat{n}}(t)%
\right],  \label{vvv12}
\end{equation}
\begin{equation}
{B}_{\hat{n}}=\sqrt{{\hat{n}}+{\delta }^{2}},  \label{vvv13}
\end{equation}
\begin{equation}
\delta =(q-ib)/2,\quad q\equiv \left( \omega _{0}-\omega \right) /g,\quad
b\equiv \gamma /g.  \label{vvv14}
\end{equation}%
As will be shown later, in realistic cases we need only the first
three terms of the expansion (\ref{ttt11}), whose constituents are
found to be
\begin{eqnarray}
\hat{\theta}_{0} &=&-ie^{-\gamma t\left( \overline{n}+1/2\right) -i\omega
\left( \hat{n}+1/2\right) t}S_{\hat{n}+1}(t)\hat{a}   \\
\hat{\theta}_{1} &=&e^{-\gamma t\left( \overline{n}+1/2\right) -i\omega \hat{%
n}t}\chi _{\hat{n}+1}(t-t_1)\chi _{\hat{n}}(-t_1)   \\
\hat{\theta}_{2} &=&-ie^{-\gamma t\left( \overline{n}+1/2\right) -i\omega (%
\hat{n}-1/2)t}\chi _{\hat{n}+1}(t-t_1)   \\
&&\times S_{\hat{n}}\left( t_1-t_2\right) \chi _{\hat{n}-1}(-t_2)\hat{a}%
^{\dagger }.  \nonumber
\end{eqnarray}

Substituting these expressions into Eqs. (\ref{ttt11}) -
(\ref{ccc}), the transition superoperator turns out to be
time-dependent, contrary to the definition of a QJS. We proceed as
in \cite{QJS}, defining the QJS as the time average of the
transition superoperator over the time $T$ (to be determined
later) during which the photoelectron is emitted with high
probability:
\begin{equation}
\hat{J}\rho \equiv \frac{1}{T}\int_{0}^{T}dt\ \hat{\Xi}(t)\rho .
\label{J-Xi}
\end{equation}%
Considering the weak coupling, $\omega ,\omega _{0}\gg \gamma
,|g|$, under which both the Jaynes--Cummings Hamiltonian
(\ref{JCH}) and the master equation (\ref{eqmestra}) are valid,
and expressing the field density operator in Fock basis as
\begin{equation}
{\rho }=\sum_{m,n=0}^{\infty }\rho _{mn}|m\rangle \langle n|,
\end{equation}%
after the averaging in (\ref{J-Xi}) the off-diagonal elements of $\hat{J}%
\rho $ vanish due to rapid oscillations of the terms $\exp (\pm
i\omega t)$. This means that the photodetection destroys the
coherence of the density matrix. This can be understood from the
point of view of information theory: information flows from the
field--sensor system to the AM, so decoherence is active;
moreover, since counting photons informs only about diagonal
elements, which are proportional to the number of photons,
non-diagonal elements can be completely ignored. Therefore, from
now on we shall treat only the diagonal elements of $\hat{J}\rho $
in (\ref{J-Xi}). Applying the superoperators $\hat{\Xi}_{l}$ on
the density matrix as in (\ref{ttt11}), after evaluating Eq.
(\ref{J-Xi}) one is left with
\begin{eqnarray}\label{rho-f}
\hat{J}\rho &=&\sum_{n=0}^{\infty }\rho
_{nn}\left[nJ_{n}^{(B)}|n-1\rangle
\langle n-1|\right. \\
&&\left.+J_{n}^{(D)}|n\rangle \langle
n|+(n+1)J_{n}^{(E)}|n+1\rangle \langle n+1|+\cdots
\right].\nonumber
\end{eqnarray}

After some lengthy however straightforward calculations one
obtains the following expression for the first term in
(\ref{rho-f})
\begin{equation}
J_{n}^{(B)} =\frac{bg\left( 1+\bar{n}\right) }{\tau }\frac{F_{n}-G_{n}}{%
\left| B_{n}\right| ^{2}},  \label{rrr11}
\end{equation}
\begin{eqnarray}
G_{n} &=&\frac{1}{(1+2\overline{n})^{2}+\phi _{n}^{2}}\Big\{1+2\overline{n}%
-\exp \left[ -\tau (1+2\overline{n})\right]  \nonumber \\
&\times &\left[ (1+2\overline{n})\cos (\tau \phi _{n})-\phi _{n}\sin (\tau
\phi _{n})\right] \Big\},  \label{rrr12}
\end{eqnarray}
\begin{equation}
F_{n} =\left\{
\begin{array}{c}
\displaystyle{\frac{\tau }{2}+\frac{1-\exp \left[ -2\tau (1+2\overline{n})%
\right] }{4(1+2\overline{n})}},\mbox{ if }\xi _{n}=\pm (1+2\overline{n}) \\
G_{n}\left( \phi _{n}\rightarrow i\xi _{n}\right), \qquad \text{otherwise},%
\end{array}%
\right.   \label{rrr13}
\end{equation}
\begin{equation}
\tau\equiv \gamma T,\qquad \phi _{n} \equiv\frac{2\mbox{Re}\left(
B_{n}\right) }{b},\qquad \text{{}}\xi _{n}\equiv%
\frac{2\mbox{Im}\left( B_{n}\right) }{b}.  \label{rrr14}
\end{equation}%
Since expression (\ref{rrr11}) is too involved to be interpreted
analytically, we shall treat it numerically, as well as ongoing
terms. In order to evaluate them we expand the time-dependent
functions ($C_n$ and $S_n$, Eq. (\ref{ggg})) in terms of
exponentials and integrate the
resulting expressions, obtaining for the second term in (\ref{rho-f}) (here for complex $%
B_{n}$ defined by Eq. (\ref{vvv13}))
\begin{equation}
J_{n}^{(D)}=\frac{g\bar{n}b^{2}\left( 1+\bar{n}\right) }{\tau }%
\sum_{j,k=1}^{4}\frac{W_{j}W_{k}^{\ast }}{y_{j}-y_{k}^{\ast }}%
\sum_{l=1}^{2}\left( -1\right) ^{l}\frac{1-e^{i\alpha _{l}\tau }}{\alpha _{l}%
},
\end{equation}%
where
\begin{eqnarray*}
&\alpha _{1}&=i(1+2\overline{n})+\left( \omega _{j}-\omega _{k}^{\ast
}\right) /b, \\
&\alpha _{2}&=\alpha _{1}+\left( y_{j}-y_{k}^{\ast }\right) /b, \\
&W_{1(2)}&=(1-\delta /B_{n+1})(1-(+)\delta /B_{n}), \\
&W_{3(4)}&=W_{1(2)} \left( B_{n+1}\rightarrow -B_{n+1}\right) , \\
&w_{1}&=w_{2}=-w_{3}=-w_{4}=B_{n+1}, \\
&y_{1}&=-y_{4}=-(B_{n+1}+B_{n}), \\
&y_{2}&=-y_{3}=B_{n}-B_{n+1}.
\end{eqnarray*}
In the same manner one can obtain exactly all the further terms,
but we shall not write the resulting expressions here.

The QJS (\ref{rho-f}) contains an infinite number of terms, so
every time the detector emits a click, the field state $\rho $
reduces to an incoherent mixture (due to the decoherence process
described above) of different states, each one with its respective
probability. Let us examine these terms more closely. The first
term, with coefficient $J_{n}^{(B)},$ takes out a photon from the
field and modifies the relative weight of the state components, so
it represents a click preceded by a photoabsorption -- we call
this event a `bright count'. The second term, dependent on
$J_{n}^{(D)}$ and proportional to $\bar{n}$ (quite small as will
be shown below), does not subtract photons from the field but only
modifies the relative weight of the state components -- it
represents the dark count, when the detector emits a click due to
the amplification of its intrinsic excitations. All
further terms in Eq. (\ref{rho-f}) are proportional to $\bar{n}^l$, $l\ge 2$%
; they describe emissions of several photons into the field upon a
click, so we call the first of these term $J_{n}^{(E)}$ the
`emission term'. We would like to stress again that all these
processes happen simultaneously with different probabilities upon
a click of the detector, so the post-measurement state of the
field is a classical mixture of these events. We also note that
there
are many different phenomena that give rise to dark counts \cite%
{Karve,Kitay,Panda}; our model takes into account only those of
them that cause the transition of the sensor from the ground to
the excited state, and since the sensor state depends on the
sensor-field interaction, the dark counts modify indirectly the
relative weight between the field components, thus depending on
$n$. The other physical phenomena that do not cause this
transition should not in principle modify the relation between the
field components and should be expressed by a constant term equal
to the corresponding dark counting rate. We shall not consider
these effects since we assume that all the intrinsic excitations
occur within the sensor.

\section{Experiments and QJS Engineering}

Now we shall compare the predictions of our model with the available
experimental data. Experimentally, the dependences of both bright
and dark counting rates are set as functions of the wavelength of
the light and the `bias parameter' (BP) of the detector. By BP we
englobe such quantities as bias voltage, bias current or any other
physical parameter the experimentalist adjusts in order to achieve
simultaneously the best Signal-to-Noise ratio $\mathcal{S}$ and the
highest bright counting rate. $\mathcal{S}$ is the ratio between the
bright counting rate $\mathcal{R}_B$ and the dark one
$\mathcal{R}_D$,
${\mathcal{S}}\equiv{\mathcal{R}_B}/{\mathcal{R}_D}$, and it has the
following useful property: as one increases the value of BP, the
bright counting rates increases while the $\mathcal{S}$ remains the
same until the \emph{breakdown} value of BP, after which the
$\mathcal{S}$ starts to fall rapidly as function of BP. So most
detectors usually operate near the breakdown BP in order to achieve
the optimal performance. Experimentally \cite{Verevkin},
$\mathcal{R}_B$ is determined by directing laser pulses containing
single photons at a given repetition rate on the detector and
calculating the rate of counts, so in our model it is described by
the term $J_{1}^{(B)}$; analogously, $\mathcal{R}_D$ is calculated
in the absence of any input signal, so it is given by $J_{0}^{(D)}$.
\begin{figure}[t]
\begin{center}
\includegraphics[width=.48\textwidth]{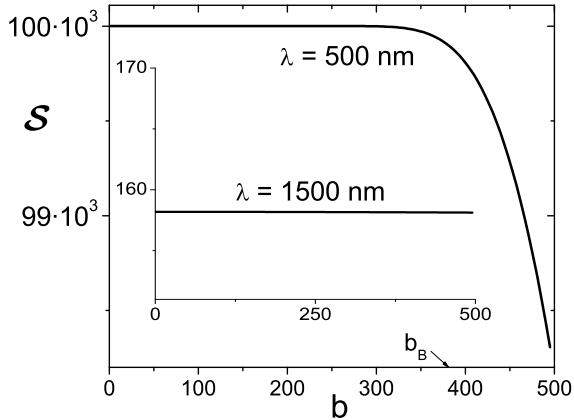} {}
\end{center}
\caption{Signal-to-Noise ratio as function of $b$ at resonance ($\protect%
\lambda=500$ nm) and far from it ($\protect\lambda=1500$ nm) in the
inset. The estimated breakdown value is $b_B\simeq 380$.}
\label{fig1}
\end{figure}

To do the comparison with experimental results first we need to set the
values of our model free parameters: $\omega _{0}$, $g$, $\tau $ and $\bar{%
n}$; for the sake of better comparison we shall express the frequencies $%
\omega _{0}$ and $\omega $ in terms of respective wavelengths $\lambda_{0}$
and $\lambda$. Thus we are left with only two experimental variables: $%
\lambda$ and $b$, where $b$ plays the role of BP. Meanwhile, our
general model can't take into account the $\mathrm{BP}\times b$
dependence for every kind of detector; nevertheless, one can argue
that BP and $b$ must be proportional to each other, since for zero
BP one should also have $b=0$, since in this case the detector
would be turned off. Fortunately, we do not need to know the exact
dependence of BP on $b$ provided we determine the breakdown value
$b_B$ corresponding to the breakdown BP at the resonance and take
it as a measure of $b$. So $b_{B}$ will be our last fixed
parameter, even though dependent on the free parameters, needed to
compare the model predictions with experiment.

After several numerical simulations we have chosen the following
values for our model free parameters that reproduce qualitatively
the common experimental behavior \cite{APD1,Verevkin,Kitay,SSPD} and
lie within the applicability region of the Jaynes--Cummings
Hamiltonian (\ref{JCH}) and the master equation (\ref{eqmestra}):
$\lambda_0=500$ nm, $g=10^{11}$ Hz, $\tau =5\times 10^{5}$ and
$\bar{n} =10^{-11}$, so $b_{B}\simeq 380$, as shown in figure 1.
Notice that 1) the values of ${{\mathcal{S}}\simeq 10^{5} }$ at the
resonance ($\lambda=500$ nm, $q=0$) and $10^{2}$ `far away' from it ($%
\lambda=1.5$ $\mu$m, $q\gg 1$) are quite realistic, and 2) the
chosen mean number of intrinsic excitations $\bar{n}$ is much larger
than the number of thermal photons as calculated from Planck's
distribution for room temperature ($\bar{n}_P\simeq 10^{-30}$),
meaning that the contribution from defects within the detector is
much stronger than the one from thermal photons (remember that we
are considering a specific part of dark counts, as described in
section II). Moreover, we verified that below $b_B$ both
$\mathcal{R}_B$ and $\mathcal{R}_D$ depend approximately linearly on
$b$, which is in agreement with our qualitative arguments.

\begin{figure}[t]
\begin{center}
\includegraphics[width=.48\textwidth]{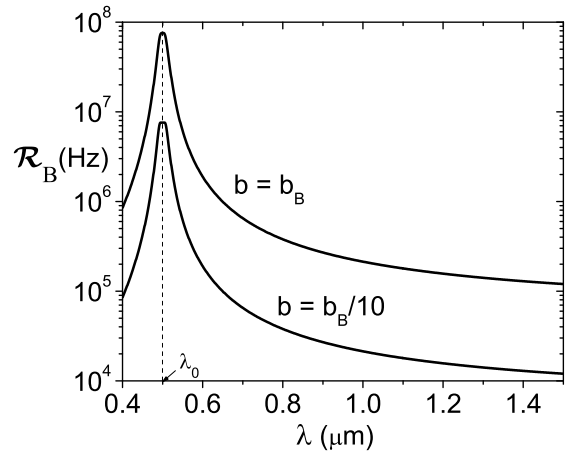} {}
\end{center}
\caption{Bright counting rate as function as wavelength of the field
for different values of $b$, showing that $\mathcal{R}_B$ increases
proportionally to $b$. The resonant wavelength is
$\protect\lambda_0=500$ nm.} \label{fig2}
\end{figure}

In figure 2 we have a plot of the $\mathcal{R}_B$ for two different
values of $b$ as function of the field wavelength, which shows a
good agreement, qualitatively and quantitatively, with experimental
results and illustrates the fact that $\mathcal{R}_B$ increases
proportionally to $b$. We could confirm numerically that
$\mathcal{R}_D$ does not depend on the field wavelength as expected,
because the dark counts are detector's internal events.

So our model agrees qualitatively in all aspects with the
experimental data. Now we turn to our main goal: how the QJS
depends on the experimental detector parameters?
First, we check that for the chosen parameters the \emph{emission terms} ($%
J_n^{(E)}$ and further terms in Eq. (\ref{rho-f})) \emph{are at
least 10 orders of magnitude smaller than the dark counts term and
even more for bright counts term}, which confirms that detectors
indeed do not emit photons into the field. Intuitively, the photon
emissions by the detector would be possible only at temperatures
much higher than room temperature through black body radiation,
which is not the case in experiments.

\begin{figure}[t]
\begin{center}
\includegraphics[width=.48\textwidth]{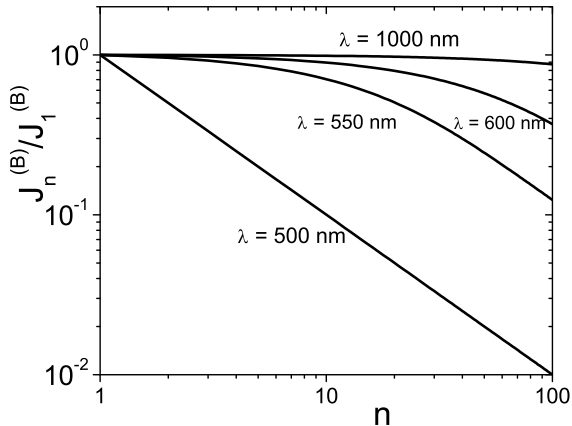} {}
\end{center}
\caption{Normalized bright counts term as function of $n$ in di-log scale at
breakdown $b_B$ for different field wavelengths. At resonance ($\protect%
\lambda=500$ nm) we have $\protect\beta\simeq 1/2$ and far away from
resonance ($\protect\lambda=1000$ nm) $\protect\beta\simeq 0$.}
\label{fig3}
\end{figure}

Thus, in practice one is dealing only with bright and dark counts
terms that act on the field simultaneously every time a count is
registered, so the QJS has the following (diagonal) form in Fock
basis
\begin{equation}
\hat{J}\rho =\mathrm{diag}\left[ \left( \hat{J}_{B}+\hat{J}_{D}\right) \rho %
\right] .
\end{equation}
In figure 3 we show the dependence of the normalized bright counts term $%
J_{n}^{(B)}/J_{1}^{(B)}$ on $n$ in di-log scale (for better visualization we
joined the points). We note that near and far away from resonance one has
nearly polynomial dependence (linear in di-log scale)
\begin{equation}
J_{n}^{(B)}\approx J_{1}^{(B)}n^{-2\beta }={\mathcal{R}_B}\
n^{-2\beta }
\end{equation}%
with $\beta \simeq 1/2$ at resonance and $\beta \simeq 0$ far away
from it. Thus, in these cases one can write the operator
dependence of bright counts term as
\begin{equation}
\hat{J}_{B}\rho ={\mathcal{R}_B}\ \left( \hat{n}+1\right) ^{-\beta
}\hat{a}\rho \hat{a}^{\dagger }\left( \hat{n}+1\right) ^{-\beta },
\end{equation}%
thus recovering the E-model with $\beta \simeq 1/2$ at the
resonance and SD-model with $\beta \simeq 0$ far away from it
($\lambda=1$ $\mu$m). This is an important result: by just
modifying the wavelength of the field one can engineer the QJS,
obtaining one of the \emph{ad hoc} proposals or another.

\begin{figure}[t]
\begin{center}
\includegraphics[width=.48\textwidth]{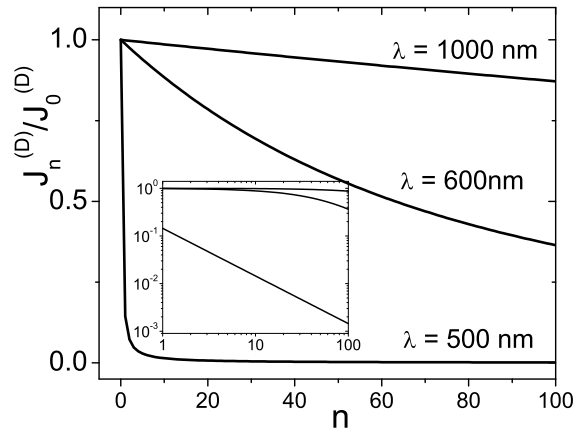} {}
\end{center}
\caption{Normalized dark counts term as function of $n$ at breakdown
$b_B$ for different field wavelengths and the same graph in di-log
scale in the inset. } \label{fig4}
\end{figure}

Now we turn to the normalized dark counts term
$J_{n}^{(D)}/J_{0}^{(D)}$, shown in figure 4 in linear scale and
in di-log scale in the inset. First, we see that out of resonance
($\lambda=1$ $\mu$m) $J_{n}^{(D)}$ is almost
independent on $n$, so we can set in this case $J_{n}^{(D)}\approx {%
\mathcal{R}_D}=\mathrm{const}$. At resonance ($\lambda=500$ nm), we
see that
for $n=0$ one gets $J_{0}^{(D)}={\mathcal{R}_D}$ and for $n>0$ we have $J_{n>0}^{(D)}\approx d\cdot{%
\mathcal{R}_D}\cdot n^{-2\beta }$, where $\beta\simeq 1/2$ and $d$
is a number less than $1$. This means that at resonance the dark
counts are suppressed by the presence of light, being generated
predominantly by vacuum fluctuations. This can be understood by the
following argument: the dark counts occur when the detector, in the
ground state, is intrinsically excited by its excitations; but at
resonance, the rate of excitations by field photons is much greater
than the one by the intrinsic excitations, so the dark counts `have
no time' to appear and therefore become suppressed. Thus, the
operator form of the dark counts term in Eq. (\ref{rho-f}) is
\begin{equation}
\hat{J}_{D}\rho ={\mathcal{R}_D}\left[ \Lambda _{0}\rho \Lambda
_{0}+d\Lambda \hat{n}^{-\beta }\rho \hat{n}^{-\beta }\Lambda \right]
,  \label{123}
\end{equation}%
where $\Lambda _{0}\equiv |0\rangle \langle 0|$, $\Lambda\equiv
1-\Lambda_0$ and at resonance we have E-model with $\beta \simeq
1/2$ and $d<1$. Far away from the resonance we recover the SD-model
with $\beta \simeq 0$ and $d=1$. Thus by operating near breakdown BP
and by varying the wavelength of the field one can engineer the QJS
and predict its bright and dark counts terms behavior; the only
inconvenience for obtaining SD-model is that the $\mathcal{S}$ ratio
is smaller than for E-model, since in this case one should operate
far away from the resonance. For the sake of completeness we could
also add to the
right-hand side of Eq. (\ref{123}) a constant term $({\mathcal{R}_D}%
^{\prime}\rho)$ for the dark counts that do not cause the transition $%
|g\rangle\rightarrow |e\rangle$ inside the sensor, however the present model
does not embraces such phenomena.

\section{Summary and conclusions}

We presented a microscopic model for a realistic photodetector in
which we modelled it as a 2-level quantum sensor plus a
macroscopic amplification mechanism. Using the quantum
trajectories approach we deduced a general QJS describing the
back-action of the detector on the field upon a photocount and
showed that it can be represented formally as an infinite sum of
terms. In that sum we have identified the terms corresponding to
the bright counts (real photoabsorptions), the dark counts and
emission events, each one occurring with its respective
probability. Adjusting the free parameters of the model to fit
experimental data, we showed that the emission terms can be
disregarded in realistic situations since their contribution
becomes insignificant, so the QJS consists effectively only of
bright and dark counts terms. Moreover, we reproduced
qualitatively and quantitatively the experimental behavior of the
counting rates and the Signal-to-Noise ratio, showing the
breakdown phenomenon. Finally, we showed that with the detector
operating near its breakdown bias one can engineer the QJS by
modifying the wavelength of the field. In particular, one recovers
the QJS's proposed previously \emph{ad hoc}: at resonance one gets
the E-model, and far away from it the SD-model is identified. Last
but not least, the contribution of the dark counts to the QJS was
derived within the context of a photocount model.

\begin{acknowledgments}
Work supported by FAPESP (SP, Brazil) contract \# 
04/13705-3. SSM and VVD acknowledge partial financial support from CNPq (DF,
Brazil). We would like to thank the anonymous referees for valuable
suggestions.
\end{acknowledgments}

\end{document}